\documentclass{LMCS}
\usepackage{amssymb,pstricks,pst-node,ded,myref,proof,url,wrapfig}
\usepackage{local}
\usepackage[show]{ed}
\input{submit.mac}
\usepackage{enumerate}

\newcounter{papercount}
{}
{}

{}
\newtheorem{Claim}[papercount]{Claim}{}

{}

{}
{}
{}

{}
{}
{}

\def\cubepts{\square_8} 

\def\deq{{\;\colon\kern-.5em=\;}}

\def\pubjsla{\cite{BenBroKoh:mehol02}}
\def\pubjslb#1{\cite{BenBroKoh:mehol02}(#1)}
\def\pubsekia{\cite{BenBroKoh:_seman_techn_higher_order_cut_elimin}}
\def\pubsekib#1{\cite{BenBroKoh:_seman_techn_higher_order_cut_elimin}(#1)}

\def\shortskip{}
\def\shortskipb{}

\def\doi{5 (1:6) 2009}
\lmcsheading%
{\doi}
{1--21}
{}
{}
{Jan.~\phantom{0}7, 2007}
{Mar.~\phantom{0}3, 2009}
{}   

\begin{document}
\title[Cut-Simulation and Impredicativity]{Cut-Simulation and 
  Impredicativity\rsuper*}

\author[C.~Benzm\"uller]{Christoph Benzm\"uller\rsuper a}
\address{{\lsuper a}Computer Laboratory, The University of Cambridge, Cambridge, CB3 0FD, England, UK}
\email{c.benzmueller@googlemail.com}
\thanks{{\lsuper{a,b}}This work has been partly funded by the SFB 378 project OMEGA at Saarland University and the EPSRC project LEO-II (grant EP/D070511/1) at Cambridge University.}

\author[C.~E.~Brown]{Chad E. Brown\rsuper b}
\address{{\lsuper b}FR 6.2 Informatik, Saarland University, D-66041 Saarbr\"ucken, Germany}
\email{cebrown@ps.uni-sb.de}

\author[M.~Kohlhase]{Michael Kohlhase\rsuper c}
\address{{\lsuper c}Computer Science, Jacobs University Bremen, D-28759 Bremen, Germany}
\email{m.kohlhase@iu-bremen.de}

\begin{abstract}
  We investigate cut-elimination and cut-simulation in impredicative (higher-order)
  logics. We illustrate that adding simple axioms such as Leibniz equations
  to a calculus for an impredicative
  logic --- in our case a sequent calculus for classical type theory --- is like adding
  cut. The phenomenon equally applies to prominent axioms like Boolean- and functional
  extensionality, induction, choice, and description. This calls for the development of
  calculi where these principles are built-in instead of being treated axiomatically.
\end{abstract}

\titlecomment{{\lsuper*}This is an extended version of the paper 'Cut-simulation in impredicative logics' presentet at IJCAR 2006.}
\maketitle

\section{Introduction}
One of the key questions of automated reasoning is the following: ``When does a set $\Phi$
of sentences have a model?'' In fact, given reasonable assumptions about calculi, most
inference problems can be reduced to determining (un)-satisfiability of a set $\Phi$ of
sentences. Since building models for $\Phi$ is hard in practice, much research in
computational logic has concentrated on finding sufficient conditions for satisfiability,
e.g. whether there is a Hintikka set $\cH$ extending $\Phi$.

Of course in general the answer to the satisfiability question depends on the class of
models at hand.  In classical first-order logic, model classes are well-understood.  In
impredicative higher-order logic, there is a whole landscape of plausible model classes
differing in their treatment of functional and Boolean extensionality.  Satisfiability
then strongly depends on these classes, for instance, the set $\Phi\deq\{a,b,qa,\lnot
qb\}$ is unsatisfiable in a model class where the universes of Booleans are required to
have at most two members (see property $\propb$ below), but satisfiable in the class
without this restriction.

In {\pubjsla} we have shown that certain (i.e. {\emph{saturated}}) Hintikka sets always
have models and have derived syntactical conditions (so-called {\emph{saturated}} abstract
consistency properties) for satisfiability from this fact. The importance of abstract
consistency properties is that one can check completeness for a calculus $\Cc$ by
verifying proof-theoretic conditions (checking that $\Cc$-irrefutable sets of formulae
have the saturated abstract consistency property) instead of performing model-theoretic
analysis (for historical background of the abstract consistency method in first-order
logic, cf.~\cite{Hintikka55,Smullyan63,Smullyan68}). Unfortunately, the saturation
condition (if $\Phi$ is abstractly consistent, then for all sentences $\bA$ one of
$\Phi\cup\{\bA\}$ or $\Phi\cup\{\neg\bA\}$ is as well) is very difficult to prove for
machine-oriented calculi (indeed as hard as cut elimination as we will show).

In this paper we investigate further the relation between the lack of
the subformula property in the saturation condition (we need to
``guess'' whether to extend $\Phi$ by $\bA$ or $\neg\bA$ on our way to
a Hintikka set) and the cut rule (where we have to ``guess,''
i.e. ``search for'' in an automated reasoning setting the cut formula
$\bA$). An important result is the insight that there exist
``cut-strong'' formulae which support the effective simulation of cut
in calculi for impredicative logics. 
Prominent examples of cut-strong
formulae are Leibniz equations and the axioms for comprehension,
extensionality, induction, description and choice. The naive addition
of any of these cut-strong formulae to any calculus for an
impredicative logic is a strong threat for effective automated proof search,
since these formulae in a way introduce the cut rule through the
backdoor (even if the original calculus is cut-free and thus appears
appropriate for proof automation at first sight). Cut-strong formulae
thus introduce additional sources for breaking the subformula property
and therefore they should either be avoided completely or treated with
great care in calculi designed for automated proof search.

Consider the following formula of higher-order logic representing Boolean extensionality:
\[\alldot{A\abtypebool}\alldot{B\abtypebool} (A\follof B)\implies
A\Leibeq^\typebool B\]
For a theorem prover to make use of this formula, it must instantiate $A$ and $B$ with terms of type $\typebool$.
In other words, the theorem prover must synthesize two arbitrary formulas.  Requiring a theorem prover to synthesize
these formulas is just as hard (and unrealistic) as requiring a theorem prover to synthesize cut formulas.
An alternative to including the formula for Boolean extensionality is to include a rule
in the search procedure which allows the theorem prover to reduce proving ${\bA\Leibeq^\typebool \bB}$ to the
subgoal of proving ${\bA \follof \bB}$.  Using this rule does not require the prover to synthesize any terms.
Simply adding such a rule is not enough to obtain a complete calculus.  We will explore what additional rules are
required to obtain completeness and argue that these rules are appropriate for mechanized proof search. 

In Section~\ref{sec:hol}, we will fix notation and review the relevant results from
{\pubjsla}. We define in Section~\ref{sec:seq-calc} a basic sequent calculus and study the
correspondence between saturation in abstract consistency classes and cut-elimination. In
Section~\ref{sec:cut-simulation} we introduce the notion of ``cut-strong'' formulae and
sequents and show that they support the effective simulation of
cut. In Section~\ref{sec:prominent-cut-strong} we demonstrate that the pertinent
extensionality axioms are cut-strong.
We develop
alternative extensionality rules which do not suffer from this problem.  Further rules are
needed to ensure Henkin completeness for this calculus with extensionality.  These new
rules correspond to the acceptability conditions we propose in
Section~\ref{sec:acceptability} to ensure the existence of models and the existence of
saturated extensions of abstract consistency classes.


\section{Higher-Order Logic}\label{sec:hol}
In {\pubjsla} we have re-examined the semantics of classical higher-order logic with the
purpose of clarifying the role of extensionality.  For this we have defined eight classes
of higher-order models with respect to various combinations of Boolean extensionality and
three forms of functional extensionality.  We have also developed a methodology of
abstract consistency (by providing the necessary model existence theorems) needed for
instance, to analyze completeness of higher-order calculi with respect to these model
classes. We now briefly summarize the main notions and results of {\pubjsla} as required
for this paper. Our impredicative logic of choice is Church's classical type theory.

\subsection{Syntax: Church's Simply Typed $\lambda$-Calculus.}
As in~\cite{Church40}, we formulate higher-order logic ($\HOL$) based on the simply typed
$\lambda$-calculus.  The set of simple types $\Types$ is freely generated from basic types
$\typebool$ and $\typeind$ using the function type constructor $\ar$.

For formulae we start with a set $\Vars$ of (typed) variables (denoted by
$X_\typea, Y, Z,$ $X^1_\typeb, X^2_\typec\ldots$) and a signature $\Signat$ of (typed) constants
(denoted by $c_\typea, f_{\typea\ar\typeb},\ldots$).  We let $\Vars_\typea$
($\Signat_\typea$) denote the set of variables (constants) of type $\typea$.  The
signature $\Signat$ of constants includes the logical constants
$\neg_{\typebool\ar\typebool}$, $\lor_{\typebool\ar\typebool\ar\typebool}$ and
$\Pi^\typea_{(\typea\ar\typebool)\ar\typebool}$ for each type $\typea$; all other
constants in $\Signat$ are called parameters.  As in {\pubjsla}, we assume there is an
infinite cardinal $\sigcard$ such that the cardinality of $\Signat_\typea$ is $\sigcard$
for each type $\typea$ (cf.~\pubjslb{3.16}).  The set of $\HOL$-formulae (or terms) are
constructed from typed variables and constants using application and
$\lambda$-abstraction.  We let $\Wff{\typea}$ be the set of all terms of type $\typea$ and
$\Wffall$ be the set of all terms.

We use vector notation to abbreviate $k$-fold applications and abstractions as
$\bA\ov{\bU^k}$ and $\lamdot{\ov{X^k}}\bA$, respectively.  We also use Church's dot
notation so that $\sdot$ stands for a (missing) left bracket whose mate is as far to the
right as possible (consistent with given brackets).  We use infix notation $\bA\lor\bB$
for $((\lor\bA)\bB)$ and binder notation $\alldot{X_\typea}\bA$ for
$(\Pi^\typea(\lamdot{X_\typea}\bA_\typebool))$.  We further use $\bA\land\bB$,
$\bA\implies\bB$, $\bA\follof\bB$ and $\exdot{X_\typea}\bA$ as shorthand for formulae
defined in terms of $\neg$, $\lor$ and $\Pi^\typea$ (cf.~{\pubjsla}).  Finally, we let
$(\bA_\typea\Leibeq^\typea\bB_\typea)$ denote the Leibniz equation
$\alldot{P\abtype{\typea\ar\typebool}}(P\bA)\implies\sdot{P\bB}$.

Each occurrence of a variable in a term is either bound by a $\lambda$
or free.  We use $\free(\bA)$ to denote the set of free variables
of $\bA$ (i.e., variables with a free occurrence in $\bA$).
We consider two terms to be equal if the terms are the same up
to the names of bound variables (i.e., we consider $\alpha$-conversion
implicitly).  A term $\bA$ is closed if $\free(\bA)$ is empty.
We let $\Wffcl\typea$ denote the set of closed terms of type $\typea$
and $\Wffclall$ denote the set of all closed terms.
Each term $\bA\in\Wffbool$ is called a proposition
and each term $\bA\in\Wffclbool$ is called a sentence.

We denote substitution of a term $\bA_\typea$ for
a variable $X_\typea$ in a term $\bB_\typeb$ by $[\bA/X]\bB$.
Since we consider $\alpha$-conversion implicitly, we assume
the bound variables of $\bB$ avoid variable capture.

Two common relations on terms are given by $\beta$-reduction and
$\eta$-reduction.  
A $\beta$-redex $(\lamdot{X}\bA)\bB$
$\beta$-reduces to $[\bB/X]\bA$.
An $\eta$-redex $(\lamdot{X}\bC X)$ (where $X\notin\free(\bC)$)
$\eta$-reduces to $\bC$.
For $\bA,\bB\in\Wffa$, we write $\bA\eqb\bB$ 
to mean $\bA$ can be converted to $\bB$ by a series of $\beta$-reductions
and expansions.
Similarly, $\bA\eqbe\bB$ means $\bA$ can be converted to $\bB$ using
both $\beta$ and $\eta$.
For each $\bA\in\Wffall$ there is a 
unique $\beta$-normal form (denoted $\Bnormform{\bA}$)
and a unique $\beta\eta$-normal form (denoted $\Benormform{\bA}$).
From this fact we know $\bA\eqb\bB$ ($\bA\eqbe\bB$) iff
$\Bnormform\bA\Metaeq\Bnormform\bB$
($\Benormform\bA\Metaeq\Benormform\bB$).

A non-atomic formula in $\Wffbool$ is any formula whose $\beta$-normal form 
is of the form $[c\ov{\bA^n}]$ where $c$ is a logical constant.
An atomic formula is any other formula in $\Wffbool$.

\subsection{Semantics: Eight Model Classes.}
A model of $\HOL$ is given by four objects: a typed collection of nonempty sets
$(\cD_\typea)_{\typea\in\Types}$, an application operator
$\appo\colon\cD_{\typea\ar\typeb}\times\cD_\typea\to\cD_\typeb$, an evaluation function
$\cE$ for terms and a valuation function
$\semival\colon\cD_\typebool\to\{\semtrue,\semfalse\}$.  A pair $(\cD,\appo)$ is called a
$\Signat$-applicative structure (cf.~\pubjslb{3.1}).  If $\cE$ is an evaluation function
for $(\cD,\appo)$ (cf.~\pubjslb{3.18}), then we call the triple $(\cD,\appo,\cE)$ a
$\Signat$-evaluation.  If $\semival$ satisfies appropriate properties, then we call the
tuple $(\cD,\appo,\cE,\semival)$ a $\Signat$-model (cf.~\pubjslb{3.40 and~3.41}).

Given an applicative structure $(\cD,\appo)$, an assignment $\phi$ is
a (typed) function from $\Vars$ to $\cD$.  An evaluation function $\cE$ maps an
assignment $\phi$ and a term $\bA_\typea\in\Wff\typea$ to an element $\cE_\phi(\bA)\in\cD_\typea$.
Evaluations $\cE$ are required to satisfy four properties
(cf.~\pubjslb{3.18}):
\begin{enumerate}[(1)]
\item $\restrict{\cE_\phi}\Vars\Metaeq\phi$.\shortskip
\item $\cE_\phi(\bF\bA)\Metaeq\cE_\phi(\bF)\appo\cE_\phi(\bA)$ for any
  $\bF\in\Wff{\typea\ar\typeb}$, $\bA\in\Wff\typea$ and types $\typea$ and
  $\typeb$.\shortskip
\item $\cE_\phi(\bA)\Metaeq\cE_\psi(\bA)$ for any type $\typea$ and $\bA\in\Wff\typea$,
  whenever $\phi$ and $\psi$ coincide on $\free(\bA)$.\shortskip
\item $\cE_\phi(\bA)\Metaeq\cE_\phi(\Bnormform\bA)$ for all $\bA\in\Wff\typea$.\shortskip
\end{enumerate}
If $\bA$ is closed, then we can simply write $\cE(\bA)$ since
the value $\cE_\phi(\bA)$ cannot depend on $\phi$.

Given an evaluation $(\cD,\appo,\cE)$, we define 
several properties a function $\semival\colon\cD_\typebool\to\{\semtrue,\semfalse\}$ may
satisfy (cf.~\pubjslb{3.40}). 

\begin{center}
 \begin{tabular}{|c|l|lcl|l|}\hline
    prop. & where & holds when & & &  for all\\\hline\hline
    $\valnot\semn$ &  $\semn\in\cD_{\typebool\ar\typebool}$ 
      & $\semival(\semn\appo\sema)\Metaeq\semtrue$ 
    &iff&  $\semival(\sema)\Metaeq\semfalse$ 
      & $\sema\in\cD_\typebool$\\\hline
    $\valor\semd$ &  $\semd\in\cD_{\typebool\ar\typebool\ar\typebool}$ 
      & $\semival(\semd\appo\sema\appo\semb)\Metaeq\semtrue$ 
    &iff& $\semival(\sema)\Metaeq\semtrue$ or $\semival(\semb)\Metaeq\semtrue$ 
      & $\sema,\semb\in\cD_\typebool$\\\hline
    $\valpi\typea\pi$ &  $\pi\in\cD_{(\typea\ar\typebool)\ar\typebool}$ 
      & $\semival(\pi\appo\semf)\Metaeq\semtrue$
    &iff& $\forall{\sema\in\cD_\typea}$ $\semival(\semf\appo\sema)\Metaeq\semtrue$
      & $\semf\in\cD_{\typea\ar\typebool}$\\\hline
    $\valeq\typea\Domeq$ & $\Domeq\in\cD_{\typea\ar\typea\ar\typebool}$ 
      & $\semival(\Domeq\appo\sema\appo\semb)\Metaeq\semtrue$ 
    &iff& $\sema\Metaeq\semb$
      & $\sema,\semb\in\cD_\typea$\\\hline
  \end{tabular}
\end{center}

\noindent A valuation
$\semival\colon\cD_\typebool\to\{\semtrue,\semfalse\}$ is required to
satisfy $\valnot{\cE(\neg)}$, $\valor{\cE(\lor)}$ and
$\valpi\typea{\cE(\Pi^\typea)}$ for every type $\typea$.

Given a model $\cM\deq (\cD,\appo,\cE,\semival)$, 
an assignment $\phi$ and
a proposition $\bA$ (or set of propositions $\Phi$),
we say $\cM$ satisfies $\bA$ (or $\Phi$) and
write $\cM\models_\phi\bA$ (or $\cM\models_\phi\Phi$)
if $\semival(\cE_\phi(\bA))\Metaeq\semtrue$
(or $\semival(\cE_\phi(\bA))\Metaeq\semtrue$ for each $\bA\in\Phi$).
If $\bA$ is closed (or every member of $\Phi$ is closed),
then we simply write $\cM\models\bA$ (or $\cM\models\Phi$)
and say $\cM$ is a model of $\bA$ (or $\Phi$).

In order to define model classes $\MODALL$ which correspond
to different notions of extensionality, we define five properties
of models (cf.~\pubjslb{3.46, 3.21 and~3.5}).
Let $\cM\deq (\cD,\appo,\cE,\semival)$ be a model. We define:\shortskip
\begin{description}
\item[$\propq$] iff for all $\typea\in\Types$ there is a
  $\Domeq^\typea\in\cD_{\typea\ar\typea\ar\typebool}$ with
  $\valeq\typea{\Domeq^\typea}$.\shortskip
\item[$\propeta$] iff $(\cD,\appo,\cE)$ is $\eta$-functional (i.e., for each $\bA\in\Wffa$
  and assignment $\phi$, $\cE_\phi(\bA)\Metaeq\cE_\phi(\Benormform{\bA})$).\shortskip
\item[$\propxi$] iff $(\cD,\appo,\cE)$ is $\xi$-functional (i.e., for each
  $\bM,\bN\in\Wff{\typeb}$, $X\in\Vars_\typea$ and assignment $\phi$,
  $\cE_\phi(\lamdot{X_\typea}\bM_\typeb)\Metaeq\cE_\phi(\lamdot{X_\typea}\bN_\typeb)$
  whenever $\cE_{\phi,[\sema/X]}(\bM)\Metaeq\cE_{\phi,[\sema/X]}(\bN)$ for every
  $\sema\in\cD_\typea$).\shortskip
\item[$\propf$] iff $(\cD,\appo)$ is functional (i.e., for each
  $\semf,\semg\in\cD_{\typea\ar\typeb}$, $\semf\Metaeq\semg$ whenever
  $\semf\appo\sema\Metaeq\semg\appo\sema$ for every $\sema\in\cD_\typea$).\shortskip
\item[$\propb$] iff $\cD_\typebool$ has at most two elements.\shortskip
\end{description}

For each $*\in\{\PD,\PETA,\PXI,\PF,\PB,\PETAB,\PXIB,\PFB\}$ (the latter set
will be abbreviated by $\cubepts$ in the remainder)
we define $\MODALL$ to be the class of all $\Signat$-models $\cM$ such that
$\cM$ satisfies property $\propq$ and each of the additional
properties $\{\propeta,\propxi,\propf,\propb\}$ indicated in
the subscript $*$ (cf.~\pubjslb{3.49}).
We always include $\beta$ in the subscript
to indicate that $\beta$-equal terms are always interpreted as
identical elements.  
We do not include property $\propq$ as an explicit subscript;
$\propq$ is treated as a basic, implicit requirement for all model classes.
See \pubjslb{3.52}
for a discussion on why we require property $\propq$.
Since we are varying four properties, one would expect to
obtain 16 model classes.  However, we showed 
in~{\pubjsla} that $\propf$ is equivalent
to the conjunction of $\propxi$ and $\propeta$.
Hence we obtain the eight model classes depicted as a cube in Figure~{\ref{landscape}}.
There are example models constructed in~{\pubjsla} to demonstrate that each of the eight model classes is distinct.
For instance, Example 5.6 of~{\pubjsla} describes how to construct a model without $\propeta$ by attaching labels to functions. 

\begin{figure}[ht]
{
    \begin{pspicture}(-2,1.5)(6,8.5)
      \rput(2,2){\rnode{fb}{\fbox{$\MODFB$}}}
      \rput(5,4){\rnode{fsb}{\fbox{$\MODETAB$}}}
      \rput(2,4){\rnode{modxib}{\fbox{$\MODXIB$}}}
      \rput(-1,4){\rnode{fsq}{\fbox{$\MODF$}}}
      \rput(-1.8,4){\rnode{fsqi}{\hbox{}}}
      \rput(-1,6){\rnode{modxi}{\fbox{$\MODXI$}}}
      \rput(-1.8,6){\rnode{fxi}{\hbox{}}}
      \rput(2,6){\rnode{fs}{\fbox{$\MODETA$}}}
      \rput(2.7,6){\rnode{fsbe}{\hbox{}}}
      \rput(5,6){\rnode{modb}{\fbox{$\MODB$}}}
      \rput(5.8,6){\rnode{fsbi}{\hbox{}}}
      \rput(2,8){\rnode{mod}{\fbox{$\MODD$}}} 
      \rput(5,8){\rnode{fsi}{\hbox{}}}
      \psset{nodesep=3pt}
      \ncline{->}{mod}{modxi}\ncput*[npos=.3]{$\propxi$}
      \ncline{->}{modxi}{fsq}\ncput*[npos=.4]{$\propeta$}
      \ncline{->}{mod}{fs}\ncput*[npos=.4]{$\propeta$}
      \ncline{->}{modb}{fsb}\ncput*[npos=.4]{$\propeta$}
      \ncline{->}{modb}{modxib}\ncput*[npos=.3]{$\propxi$}
      \ncline{->}{mod}{fsq}\ncput*[npos=.5]{$\propf$}
      \ncline{->}{fs}{fsq}\ncput*[npos=.3]{$\propxi$}
      \ncline{->}{modb}{fb}\ncput*[npos=.5]{$\propf$}
      \ncline{->}{fs}{fsb}\ncput*[npos=.2]{$\propb$}
      \ncline{->}{modxi}{modxib}\ncput*[npos=.7]{$\propb$}
      \ncline{->}{mod}{modb}\ncput*[npos=.7]{$\propb$}
      \ncline{->}{fsq}{fb}\ncput*[npos=.7]{$\propb$}
      \ncline{->}{fsb}{fb}\ncput*[npos=.3]{$\propxi$}
      \ncline{->}{modxib}{fb}\ncput*[npos=.4]{$\propeta$}
    \end{pspicture}
} 
\caption{\label{landscape} The Landscape of $\HOL$-Semantics.}
\end{figure} 
Special cases of $\Signat$-models are Henkin models and standard models (cf.~\pubjslb{3.50 and~3.51}).
A Henkin model is a model in $\MODFB$ such that the applicative structure $(\cD,\appo)$
is a frame, i.e. $\cD_{\typea\ar\typeb}$ is a subset of the function space $(\cD_\typeb)^{\cD_\typea}$
for each $\typea,\typeb\in\Types$ and $\appo$ is function application.
A standard model is a Henkin model in which $\cD_{\typea\ar\typeb}$ is the full function space $(\cD_\typeb)^{\cD_\typea}$.
Every model in $\MODFB$ is isomorphic to a Henkin model (see the discussion
following \pubjslb{3.68}).

\subsection{Saturated Abstract Consistency Classes and Model Existence.}
Finally, we review the model existence theorems proved in~{\pubjsla}.  There are three
stages to obtaining a model in our framework.  First, we obtain an abstract consistency
class $\acc$ (usually defined as the class of irrefutable sets of sentences with respect
to some calculus).  Second, given a (sufficiently pure) set of sentences $\Phi$ in the
abstract consistency class $\acc$ we construct a Hintikka set $\cH$ extending $\Phi$.
Third, we construct a model of this Hintikka set (and hence a model of $\Phi$).

A $\Signat$-abstract consistency class $\acc$ is a class of sets of
$\Signat$-sentences.  An abstract consistency class is always required to
be closed under subsets (cf.~\pubjslb{6.1}).  Sometimes we require the
stronger property that $\acc$ is compact, i.e. a set $\Phi$ is in $\acc$
iff every finite subset of $\Phi$ is in $\acc$ (cf. \pubjslb{6.1,6.2}).

To describe further properties of abstract consistency classes, we use the
notation $S*a$ for $S\cup\{a\}$ as in~{\pubjsla}.  The following is a list
of properties a class $\acc$ of sets of sentences can satisfy with respect
to arbitrary $\Phi\in\acc$
(cf.~\pubjslb{6.5}):\shortskip
  \begin{description}
  \absitem{\absc} If $\bA$ is atomic, then $\bA\notin\Phi$ or
    $\neg\bA\notin\Phi$.\shortskip
  \absitem{\absneg} If $\neg\neg\bA\in\Phi$, then
    $\Phi*\bA\in\acc$.\shortskip
  \absitem{\absbeta} If $\bA\eqb\bB$ and $\bA\in\Phi$, then
    $\Phi*\bB\in\acc$.\shortskip
  \absitem{\absbe} If $\bA\eqbe\bB$ and $\bA\in\Phi$, then
    $\Phi*\bB\in\acc$.\shortskip
  \absitem{\absor} If $\bA\lor\bB\in\Phi$, then
    $\Phi*\bA\in\acc$ or $\Phi*\bB\in\acc$.\shortskip
  \absitem{\absand} If $\neg(\bA\lor\bB)\in\Phi$, then
    $\Phi *\neg\bA *\neg\bB\in\acc$.\shortskip
  \absitem{\absforall} If $\Pi^\typea\bF\in\Phi$, then
    $\Phi*\bF\bW\in\acc$ for each $\bW\in\cWffa$.\shortskip
  \absitem{\absexists} If $\neg\Pi^\typea\bF\in\Phi$, then
    $\Phi*\neg(\bF w)\in\acc$ for any parameter $w_\typea\in\Signat_\typea$ which
    does not occur in any sentence of $\Phi$.\shortskip
  \absitem{\absb}\label{Def:acc-b:b} If
    $\neg(\bA\Leibeq^\typebool\bB)\in\Phi$, then $\Phi *\bA *\neg\bB\in\acc$ or
    $\Phi *\neg\bA *\bB\in\acc$.\shortskip
  \absitem{\absxi}\label{Def:acc-q:xi} If
    $\neg(\lamdot{X_\typea}\bM\Leibeq^{\typea\ar\typeb}\lamdot{X_\typea}\bN)\in\Phi$,
    then $\Phi*\neg([w/X]\bM\Leibeq^\typeb[w/X]\bN)\in\acc$ for any parameter
    $w_\typea\in\Signat_\typea$ which does not occur in any sentence of $\Phi$.\shortskip
  \absitem{\absf}\label{Def:acc-q:q} If
    $\neg(\bG\Leibeq^{\typea\ar\typeb}\bH)\in\Phi$, then $\Phi*\neg(\bG
    w\Leibeq^\typeb\bH w)\in\acc$ for any parameter $w_\typea\in\Signat_\typea$
    which does not occur in any sentence of $\Phi$.\shortskip
  \absitem{\abssat} Either $\Phi *\bA\in\acc$ or $\Phi *
    \neg\bA\in\acc$.
  \end{description}

\noindent We say $\acc$ is an abstract consistency class if it is
closed under subsets and satisfies $\absc$, $\absneg$, $\absbeta$,
$\absor$, $\absand$, $\absforall$ and $\absexists$.  We let $\ACCMODD$
denote the collection of all abstract consistency classes.  For each
$*\in\cubepts$ we refine $\ACCMODD$ to a collection $\ACCstar$ where
the additional properties $\{\absbe,\absxi,\absf,\absb\}$ indicated by
$*$ are required (cf.~\pubjslb{6.7}).  We say an abstract consistency
class $\acc$ is saturated if $\abssat$ holds. 

Using $\absc$ (atomic consistency) and the fact that there are infinitely
many parameters at each type, we can show every abstract consistency class
satisfies non-atomic consistency.  That is, for every abstract consistency
class $\acc$, $\bA\in\Wffclbool$ and $\Phi\in\acc$, we have either
$\bA\notin\Phi$ or $\neg\bA\notin\Phi$ (cf.~\pubjslb{6.10}).

In \pubjslb{6.32} we show that sufficiently $\Signat$-pure sets in
saturated abstract consistency classes extend to saturated Hintikka
sets. (A set of sentences $\Phi$ is sufficiently $\Signat$-pure if for
each type $\typea$ there is a set $\cP_\typea$ of parameters of type
$\typea$ with cardinality $\sigcard$ and such that no parameter in
$\cP$ occurs in a sentence in $\Phi$. A Hintikka set is a maximal
element in an abstract consistency class.)

In the Model Existence Theorem for Saturated Sets \pubjslb{6.33} we show
that these saturated Hintikka sets can be used to construct models $\cM$
which are members of the corresponding model classes $\MOD_{*}$.  Then we
conclude (cf.~\pubjslb{6.34}):\\[-1em]

\noindent \textbf{Model Existence Theorem for Saturated Abstract Consistency Classes}: {\em{For all
    $*\in\cubepts$, if $\acc$ is a saturated abstract consistency class in $\ACC{*}$ and
    $\Phi\in\acc$ is a sufficiently $\Signat$-pure set of sentences, then there exists a
    model $\cM\in\MOD_{*}$ that satisfies $\Phi$.  Furthermore, each domain of $\cM$ has
    cardinality at most $\sigcard$.}}\\[-1em]

In \pubjsla~we apply the abstract consistency method to analyze
completeness for different natural deduction calculi.  Unfortunately, the
saturation condition is very difficult to prove for machine-oriented
calculi (indeed as we will see in Section~\ref{sec:seq-calc} it is
equivalent to cut elimination), so Theorem~\pubjslb{6.34} cannot be easily
used for this purpose directly.

In Section~\ref{sec:acceptability} we therefore motivate and present a set
of extra conditions for $\ACCMODFB$ we call {\defemph{acceptability}} conditions.
The new conditions are sufficient to prove model existence.

\section{Sequent Calculi, Cut and Saturation}\label{sec:seq-calc}
We will now study cut-elimination and cut-simulation with respect to (one-sided) sequent
calculi.

\subsection{Sequent Calculi $\SEQCALC$.}
  We consider a sequent to be a finite set $\Delta$ of $\beta$-normal
sentences from $\Wffclbool$.  A sequent calculus $\SEQCALC$ provides an
inductive definition for when $\thdash_\SEQCALC\Delta$ holds.  We say a
sequent calculus rule
{
\begin{center} 
$\icnc{\Delta_1}{\cdots}{\Delta_n}{\Delta}{r}$ 
\end{center}}
\noindent is \defin{admissible} in $\SEQCALC$ if $\thdash_\SEQCALC\Delta$ holds whenever
$\thdash_\SEQCALC\Delta_i$ for all $1\leq i\leq n$.  For any natural number $k\geq 0$, we
call an admissible rule $r$ \defin{$k$-admissible} if any instance of $r$ can be replaced
by a derivation with at most $k$ additional proof steps.  Given a sequent $\Delta$, a
model $\cM$, and a class $\MOD$ of models, we say $\Delta$ is {\em{valid for $\cM$\/}} (or
{\em{valid for $\MOD$\/}}), if $\cM\models\bD$ for some $\bD\in\Delta$ (or $\Delta$ is
valid for every $\cM\in\MOD$).  As for sets in abstract consistency classes, we use the
notation $\Delta*\bA$ to denote the set $\Delta\cup\{\bA\}$ (which is simply $\Delta$ if
$\bA\in\Delta$).  Figure~\ref{seqcalc} introduces several sequent calculus rules.  Some of
these rules will be used to define sequent calculi, while others will be shown admissible
(or even $k$-admissible).

\begin{figure}[ht]
\underline{Basic Rules} \hfill
 \ianc{\bA{\mbox{ atomic (and $\beta$-normal)}}}{\Delta*\bA*\neg\bA}{\seqinit}  \hspace{3em}
  \ianc{\Delta * \bA}
       {\Delta * \neg\neg\bA}
       {\seqneg} \\[.5em]
\phantom{Bla} \hfill 
  \ibnc{\Delta *\neg\bA}
       {\Delta *\neg\bB}
       {\Delta *\neg(\bA\lor\bB)}
       {\seqlorl}\hspace{3em}
  \ianc{\Delta *\bA *\bB}
       {\Delta * (\bA\lor\bB)}
       {\seqlorr}\\[.5em]
\phantom{Bla} \hfill 
  \ibnc{\Delta * {\neg\bnormform{(\bA\bC)}}}
       {\bC\in\Wffcl\typea}
       {\Delta *\neg\Pi^\typea\bA}
       {\seqpil}\hspace{3em}
 \ibnc{\Delta *\bnormform{(\bA c)}}
       {c_\typea\in\Signat\ new}
       {\Delta *\Pi^\typea\bA}
       {\seqpir}\\[.5em]
\underline{Inversion Rule}\hfill
  \ianc{\Delta *\neg\neg\bA}
       {\Delta *\bA}
       {\seqneginv}\\[.5em]
\underline{Weakening and Cut Rules}\hfill
     \ianc{\Delta}
           {\Delta\cup\Delta'}
           {\seqweak}\hspace{3em}
      \ibnc{\Delta*\bC}{\Delta*\neg\bC}
         {\Delta}
         {\seqcut}
\caption{\label{seqcalc} Sequent Calculus Rules}
\end{figure}

\begin{rem}[Alternative Formulations]
  There are many kinds of sequent calculi given in the literature.
  We could have chosen to work with two sided sequents.  This choice would have
  allowed us to generalize many of our results to the intuitionistic case.
  The notion of cut-strong formulae could still be defined and many of our
  examples of cut-strong formulae would also be cut-strong in the intuitionistic case.
  On the other hand, assuming we only treat the classical case, we could restrict to negation normal forms
  in the same way that we restrict to $\beta$-normal forms.
  This would eliminate the need to consider the rules $\seqneg$ and $\seqneginv$.
  Both of these alternatives are reasonable.  The choices we have made are
  for ease of presentation and to make the connection with~{\pubjsla} as simple as possible. 
\end{rem}


\subsection{Abstract Consistency Classes for Sequent Calculi.} For any sequent calculus $\SEQCALC$ we can define a class $\accseq\SEQCALC$
of sets of sentences.  Under certain assumptions, $\accseq\SEQCALC$ is an
abstract consistency class.  First we adopt the notation $\neg\Phi$ and
$\bnormform\Phi$ for the sets $\{\neg\bA |\bA\in\Phi\}$ and
$\{\bnormform\bA |\bA\in\Phi\}$, resp., where $\Phi\subseteq\cWffbool$.
Furthermore, we assume this use of $\neg$ binds more strongly than $\cup$
or $*$, so that $\neg\Phi\cup\Delta$ means $(\neg\Phi)\cup\Delta$ and
$\neg\Phi *\bA$ means $(\neg\Phi)*\bA$.

\begin{defi}\label{def:accseq}
  Let $\SEQCALC$ be a sequent calculus.
  We define $\accseq\SEQCALC$ to be the class
  of all finite $\Phi\subset\cWffbool$ such that
  $\thdash_\SEQCALC\neg\bnormform\Phi$ does not hold.
\end{defi}

In a straightforward manner, one can prove the following
results (see the Appendix).

\begin{lem}\label{lem:accseq-notin}
  Let $\SEQCALC$ be a sequent calculus such that $\seqneginv$ is
  admissible.  For any finite sets $\Phi$ and $\Delta$ of sentences,
  if $\Phi\cup\neg\Delta\notin\accseq\SEQCALC$, then
  $\thdash_\SEQCALC\neg\bnormform\Phi\cup\bnormform\Delta$ holds.
\end{lem}

\begin{thm}\label{thm:accseq-acc}
  Let $\SEQCALC$ be a sequent calculus.  If the rules $\seqneginv$,
  $\seqneg$, $\seqweak$, $\seqinit$, $\seqlorl$, $\seqlorr$, $\seqpil$ and
  $\seqpir$ are admissible in $\SEQCALC$, then
  $\accseq\SEQCALC\in\ACCMODD$.
\end{thm}

We can furthermore show the following relationship between saturation and
cut (see the Appendix).

\begin{thm}\label{lem:cut-implies-set}
  Let $\SEQCALC$ be a sequent calculus.\shortskip
  \begin{enumerate}[\em(1)]
  \item If $\seqcut$
    is admissible in $\SEQCALC$, then $\accseq\SEQCALC$ is saturated.\shortskipb
  \item If $\seqneg$ and $\seqneginv$ are admissible in $\SEQCALC$
    and $\accseq\SEQCALC$ is saturated, then $\seqcut$ is admissible in $\SEQCALC$.
  \end{enumerate}
\end{thm}

\noindent Since saturation is equivalent to admissibility of cut,
we need weaker conditions than saturation.  A natural
condition to consider is the existence of saturated extensions.

\begin{defi}[Saturated Extension]
  Let $*\in\cubepts$ and $\acc,\acc'\in\ACC{*}$ be abstract consistency classes.
  We say $\acc'$ is an {\defin{extension}} of $\acc$
  if $\Phi\in\acc'$ for every sufficiently $\Signat$-pure $\Phi\in\acc$.
  We say $\acc'$ is a {\defin{saturated extension}} of $\acc$
  if $\acc'$ is saturated and an extension of $\acc$.
\end{defi}

There exist abstract consistency classes $\Gamma$ in $\ACCMODFB$ which have
no saturated extension.

\begin{exa}\label{ex:nosatb}
  Let $a_\typebool,b_\typebool,q_{\typebool\ar\typebool}\in\Signat$ and
  $\Phi\deq\{a,b,(qa),\neg (q b)\}$. We construct an abstract consistency class $\acc$
  from $\Phi$ by first building the closure $\Phi'$ of $\Phi$ under relation $\eqb$ and
  then taking the power set of $\Phi'$.  It is easy to check that this $\acc$ is in
  $\ACCMODFB$.  Suppose we have a saturated extension $\acc'$ of $\acc$ in $\ACCMODFB$.
  Then $\Phi\in\acc'$ since $\Phi$ is finite \textup{(}hence sufficiently
  $\Signat$-pure\textup{)}.  By saturation, $\Phi*(a\Leibeq^\typebool b)\in\acc'$ or
  $\Phi*\neg(a\Leibeq^\typebool b)\in\acc'$.  In the first case, applying $\absforall$
  with the constant $q$, $\absor$ and $\absc$ contradicts $(q a),\neg (q b)\in\Phi$.  In
  the second case, $\absb$ and $\absc$ contradict $a,b\in\Phi$.
\end{exa}

Existence of any saturated extension of a sound sequent calculus $\SEQCALC$
implies admissibility of cut.  The proof uses the model existence theorem
for saturated abstract consistency classes (cf.~\pubjslb{6.34}).  The proof is in the Appendix.

\begin{thm}\label{thm:satext-implies-cut}
  Let $\SEQCALC$ be a sequent calculus
  which is sound for $\MODALL$.
  If $\accseq\SEQCALC$ has a saturated extension
  $\acc'\in\ACCstar$,
  then $\seqcut$ is admissible in $\SEQCALC$.
\end{thm}


\subsection{Sequent Calculus $\SEQCALCD$.}
We now study a particular sequent calculus $\SEQCALCD$ defined by the rules
$\seqinit$, $\seqneg$, $\seqlorl$, $\seqlorr$, $\seqpil$ and $\seqpir$ 
(cf. Figure~\ref{seqcalc}). It is easy to show that $\SEQCALCD$
is sound for the eight model classes 
and in particular
for class $\MODD$.

The reader may easily prove the following Lemma.
\begin{lem} \label{lemma-length} Let $\bA\in\cWffbool$ be an atom,
  $\bB\in\cWffa$, and $\Delta$ be a sequent. 
\begin{enumerate}[\em(1)]
\item\label{lemma-length-a} $\Delta*\bA\Leftrightarrow\bA\deq\Delta*\neg(\neg(\neg\bA
 \vee\bA)\vee\neg(\neg\bA\vee\bA))$ is derivable in $7$ steps in $\SEQCALCD$.
\item\label{lemma-length-b} $\Delta*\bB\Leibeq^\typea\bB\deq\Delta*\Pi^\typea
  (\lamdot{P_{\typea\ar\typebool}}\neg (P\bB)\vee (P\bB)$ is derivable in
  $3$ steps  in
$\SEQCALCD$.
\end{enumerate}
\end{lem} 

The proof of the next Lemma is by induction on derivations and 
is given in the Appendix. 

\begin{lem}\label{lem:admissible-rules} 
The rules $\seqneginv$ and $\seqweak$ are $0$-admissible in $\SEQCALCD$.
\end{lem}

\begin{thm}\label{thm:complete-beta}
The sequent calculus $\SEQCALCD$ is complete for the model class $\MODD$
 and the rule $\seqcut$ is admissible.
\end{thm} 

\proof  By Theorem~\ref{thm:accseq-acc} and Lemma~\ref{lem:admissible-rules},
$\accseq\SEQCALCD\in\ACCMODD$. Suppose $\thdash_{\SEQCALCD}\Delta$ does not
hold. Then $\neg\Delta\in\ACCMODD$ by Lemma~\ref{lem:accseq-notin}. 
By the model existence theorem for $\ACCMODD$ (cf. {\pubsekib{8.1}}) there
exists a model for $\neg\Delta$ in $\MODD$. This gives completeness of 
$\SEQCALCD$. We can use completeness to conclude cut is
admissible in $\SEQCALCD$.\qed

Andrews proves admissibility of cut for a sequent calculus similar to $\SEQCALCD$
in~\cite{Andrews:ritt71}.  The proof in~\cite{Andrews:ritt71} contains the essential
ingredients for showing completeness.


While $\seqcut$ is admissible in $\SEQCALCD$ the next theorem shows
that $\seqcut$ is not $k$-admissible in $\SEQCALCD$ for any
$k\in{\mathbb{N}}$, which means $\SEQCALCD$ is not only superficially
cut-free and that by adding $\seqcut$ to $\SEQCALCD$ we can
achieve significantly shorter proofs.

\begin{thm}\label{firstclaim}
  $\seqcut$ is not $k$-admissible in $\SEQCALCD$ for any $k\in{\mathbb{N}}$.
\end{thm}
\proof The proof is not formally worked out here; we only sketch the
argumentation: The main idea is to show that the hyper-exponential
speed-up results known for first-order logic do transfer to (the
first-order fragment of) our calculus.  For this, we compare our
sequent calculus $\SEQCALCD$ with a standard first-order variant of it
which we call $\SEQCALCD^{FO}$ (this only requires appropriate
modifications of the rules $\seqpil$ and $\seqpir$). 
Clearly, any first-order sequent which can be derived in
$\SEQCALCD^{FO}$ can be derived in $\SEQCALCD$ with the same number
of steps (using essentially the same derivation).  
More interestingly, one can show that for any 
derivation $\cD$ in $\SEQCALCD$ of a first-order sequent $\Delta$
there is a derivation $\cD'$ in $\SEQCALCD^{FO}$ of $\Delta$
with the same number of rule applications.  (During the induction,
one collapses higher-order terms to first-order terms
in such a way that first-order terms collapse to themselves.)
Thus no speedup with respect to first-order provability
can be achieved by using $\SEQCALCD$ instead of 
the cut-free first-order sequent calculus $\SEQCALCD^{FO}$.
Finally we refer to the
following results:
\begin{enumerate}[$\bullet$]
\item Theorem 5.2.13 in~\cite{TroelstraSchw00b} shows that for a classical first-order
  sequent calculus there is at least an exponential speed-up of proofs with cut.
  Furthermore, Propositions 6.11.3 and 6.11.4 there show a related hyper-exponential
  speed-up result.\shortskip
\item An example for hyper-exponential speed-up is also given in
  {\cite{StatmanBoundsForProof,Orevkov79}}. \nocite{Orevkov79e}\shortskip
\item In higher-order logic the speed-up should be faster than any primitive recursive
  function according to the ``curious inference'' George Boolos presents
  in~\cite{BoolosCuriousInference}.\shortskip
\end{enumerate}
\qed

We will now show that $\seqcut$ actually becomes $k$-admissible in $\SEQCALCD$ if certain
formulae are available in the sequent $\Delta$ we wish to prove.
\section{Cut-Simulation}\label{sec:cut-simulation}

\subsection{Cut-Strong Formulae and Sequents.} $k$-cut-strong formulae can be used to
effectively simulate cut. Effectively means that the elimination of each application of a
cut-rule introduces maximally $k$ additional proof steps, where $k$ is constant.

\begin{defi}
  Given an arbitrary but fixed number $k>0$.  We call formula $\bA\in\cWffbool$
  \defemph{$k$-cut-strong} for $\SEQCALC$ (or simply \defemph{cut-strong}) if the
  following cut rule variant is $k$-admissible in $\SEQCALC$:\footnote{Here, we could
    alternatively use ($k$-)derivability (see~\cite{HindleySeldin86}) to give a stronger
    but less general notion of $k$-cut-strongness. In fact, all axioms we discuss in this
    paper would remain $k$-cut-strong.  From a proof theoretic point of view one may argue
    that this alternative notion leads to a more interesting result although it may
    generally apply to fewer axioms.  }
\begin{center} 
$     \ibnc{\Delta*\bC}{\Delta*\neg\bC}
         {\Delta*\neg\bA}
         {\seqcuta}
$   
\end{center}
We can alternative state the condition for $\bA$ to be \defemph{$k$-cut-strong} for $\SEQCALC$ as follows:
   For all $\Delta$ and $\bC$, if $\thdash_\SEQCALC \Delta * \bC$ in $n$ steps and
   $\thdash_\SEQCALC \Delta * \neg\bC$ in $m$ steps, then $\thdash_\SEQCALC \Delta * \neg \bA$ in at most $n+m+k$ steps.

\end{defi}


Our examples below illustrate that cut-strength of a formula usually only
weakly depends on the calculus $\SEQCALC$: it only presumes standard
ingredients such as $\beta$-normalization, weakening, and rules for the
logical connectives.

We present some simple examples of cut-strong formulae for our sequent calculus
$\SEQCALCD$. A corresponding phenomenon is observable in other higher-order
calculi, for instance, for the calculi presented in
\cite{Andrews:ritt71,Ben99,Brown2004a,Huet:amott73}.

\begin{exa}\label{ex1} 
  The Formula $\alldot{P_\typebool} P\deq\Pi^\typebool(\lamdot{P_\typebool} P)$ is
  $3$-cut-strong in $\SEQCALCD$. This is justified by the following derivation which
  actually shows that rule $\seqcuta$ for this specific choice of $\bA$ is derivable in
  $\SEQCALCD$ by maximally $3$ additional proof steps. The only interesting proof step is
  the instantiation of $P$ with formula $\bD\deq\neg\bC\vee {\bC}$ in rule
  $\seqpilg{\bD}$. (Note that $\bC$ must be $\beta$-normal; sequents such as $\Delta*\bC$
  by definition contain only $\beta$-normal formulae.)
{
\[
  \infer[\seqpilg\bD]
        {\Delta*\neg\Pi^\typebool(\lamdot{P_\typebool} P)}
        {\infer[\seqlorl]
          {\Delta*\neg(\neg {\bC}\vee {\bC})}
          {
            \infer[\seqneg]{\Delta*\neg\neg {\bC}}{\Delta*{\bC}}
            &
            \Delta*\neg\bC
            }}
\]
}
  \unskip\noindent Clearly, $\alldot{P_\typebool} P$ is not a very
  interesting cut-strong formula since it implies falsehood,
  i.e. inconsistency.
\end{exa}
\begin{exa}\label{ex2}
  The formula $\alldot{P_\typebool} P\implies P\deq
 \Pi^\typebool(\lamdot{P_\typebool}\neg P\vee P)$ is $3$-cut-strong in $\SEQCALCD$. This is an
  example of a tautologous cut-strong formula. Now $P$ is 
  simply instantiated with $\bD\deq{\bC}$ in rule $\seqpilg{\bD}$. Except
  for this first step the derivation is identical to the one for Example~\ref{ex1}.



\end{exa}
\begin{exa}\label{ex3}
  Leibniz equations $\bM \Leibeq^\alpha \bN \deq$ $\Pi^\alpha(\lamdot{P}\neg P\bM $ $\vee\  P\bN)$ (for
  arbitrary formulae $\bM,\bN\in\cWffa$ and types $\alpha\in\Types$) are
  $3$-cut-strong in $\SEQCALCD$. This includes the special cases
  $\bM\Leibeq^\alpha\bM$. Now $P$ is instantiated with
  $\bD\deq\lamdot{X_\typea} {\bC}$  in rule $\seqpilg{\bD}$.  Except
  for this first step the derivation is identical to the one for Example~\ref{ex1}.


\end{exa}
\begin{exa}\label{ex4}
  The original formulation of higher-order logic (cf.~\cite{Russell08}) contained
  comprehension axioms of the form
  $\Cc\deq\exdot{P_{\typea^1\ar\cdots\ar\typea^n\ar\typebool}\alldot{\ov{X^n}}
    P{\ov{X^n}}\follof\bB_\typebool}$ where $\bB_\typebool\in\Wffbool$ is arbitrary with
  $P\notin\free (\bB)$.  Church eliminated the need for such axioms by formulating
  higher-order logic using typed $\lambda$-calculus. We will now show that the instance
  $\Cc^{I}\deq\exdot{P_{\typeind\ar\typebool}}\alldot{X_\typeind} P X\follof X
 \Leibeq^\typeind X $
is $16$-cut-strong in $\SEQCALCD$ (note that $\seqweak$ is
$0$-admissible). This motivates building-in comprehension principles
instead of treating comprehension axiomatically.  

{
\[
\infer[\seqneg]
  {\Delta*\Cc^I}
  {\infer[\seqpirg{p_{\typeind\ar\typebool}}]
    {\Delta*\Pi^{{\typeind\ar\typebool}} (\lamdot{P^{\typeind\ar\typebool}}\neg\Pi^\typeind(\lamdot{X_\typeind} p X\follof X
        \Leibeq^\typeind X))}
    {\infer[\seqpilg{a_\typeind}]
      {\Delta*\neg\Pi^\typeind(\lamdot{X_\typeind} p X\follof X
       \Leibeq^\typeind X)}
      {\infer[\seqneg]
        {\Delta*\neg\neg (\neg (p a\implies a\Leibeq^\typeind a)\vee\neg (a
         \Leibeq^\typeind a\implies p a)) }
        {\infer[\seqlorr]
          {\Delta*\neg (p a\implies a\Leibeq^\typeind a)\vee\neg (a
           \Leibeq^\typeind a\implies p a)}
          {\infer[\seqlorl]
            {\Delta*\neg (p a\implies a\Leibeq^\typeind a) *\neg (\neg (a
             \Leibeq^\typeind a)\vee p a)}
            {\infer[\seqneg]
                {\Delta*\neg (p a\implies a\Leibeq^\typeind a) *\neg\neg (a
                \Leibeq^\typeind a)}
                {\infer*[]
                  {\Delta*\neg (p a\implies a\Leibeq^\typeind a) * a
 \Leibeq^\typeind a}
                  {\mbox{3 steps; see Lemma~\ref{lemma-length}}}}
             &  
            \cD}
          }
        }
      }   
    }
  }
\]}
Derivation $\cD$ is:\shortskip
{
\[
\infer[\seqlorl]
  {\Delta*\neg (\neg p a\vee a\Leibeq^\typeind a) *\neg pa}
  {\infer[\seqneg]
     {\Delta*\neg\neg p a *\neg pa}
     {\infer[\seqinit]{\Delta * pa *\neg pa}{}}
   &
  \infer[\seqweak]
     {\Delta*\neg (a\Leibeq^\typeind a) *\neg pa}
     {\infer*[\mbox{3 steps; see Ex.~\ref{ex3}}]{\Delta*\neg(a\Leibeq^\typebool a)}{\deduce{}{\Delta*\bC & \Delta*\neg\bC}}}}
\]
}
\end{exa}

As we will show later, many  prominent axioms for higher-order logic
also belong to the class of cut-strong formulae.

\subsection{Cut-Simulation.}

The cut-simulation theorem is a main result of this paper. It says that cut-strong
sequents support an effective simulation (and thus elimination) of cut in
$\SEQCALCD$. Effective means that the size of cut-free derivation grows only linearly for
the number of cut rule applications to be eliminated.

\begin{defi} 
  A sequent $\Delta$ is called \defin{$k$-cut-strong} (or simply \defin{cut-strong}) if
  there exists a 
  $k$-cut-strong formula $\bA\in\cWffbool$ such that
  $\neg\bA\in\Delta$. We call $\bA$ the \defin{$k$-realizer} of $\Delta$.
\end{defi}

We first fix the following calculi: Calculus $\SEQCALCD^{cut}$ extends $\SEQCALCD$ by the
rule $\seqcut$ and calculus $\SEQCALCD^{cut^\bA}$ extends $\SEQCALCD$ by the rule
${\seqcuta}$ for some arbitrary but fixed cut-strong formula $\bA$.

\begin{thm}\label{thm:cut-simulation-a}
  Let $\Delta$ be a $k$-cut-strong sequent with realizer $\bA$. For each derivation
  $\cD\colon\thdash_{\SEQCALCD^{cut}}\Delta$ with $d$ proof steps there is an alternative
  derivation $\cD'\colon\thdash_{\SEQCALCD^{cut^\bA}}\Delta$ with $d$ proof steps.
\end{thm}

\proof Note that the rules $\seqcut$ and $\seqcuta$ coincide whenever
$\neg\bA\in\Delta$. Intuitively, we can replace each occurrence of $\seqcut$ in $\cD$ by
$\seqcuta$ in order to obtain a $\cD'$ of same size. Technically, in the induction proof
one must weaken to ensure $\neg\bA$ stays in the sequent and carry out a parameter
renaming to make sure the eigenvariable condition is satisfied.  \qed

\begin{thm}\label{thm:cut-simulation-b}
  Let $\Delta$ be a $k$-cut-strong sequent with realizer $\bA$.  For each derivation
  $\cD\colon\thdash_{\SEQCALCD^{cut^\bA}}\Delta$ with $d$ proof steps and with $n$
  applications of rule $\seqcut$ there exists an alternative derivation
  $\cD'\colon\thdash_{\SEQCALCD}\Delta$ with maximally $d+nk$ proof steps.
\end{thm}

\proof  $\bA$ is $k$-cut-strong so by definition $\seqcuta$ is
$k$-admissible in $\SEQCALCD$. This means that $\seqcuta$ can be
eliminated in $\cD$ and each single elimination of $\seqcuta$ introduces
maximally $k$ new proof steps. Now the assertion can be easily obtained by
a simple induction over $n$. \qed

\begin{cor}\label{thm:cut-simulation-c}
  Let $\Delta$ be a $k$-cut-strong sequent. For each derivation
  $\cD\colon\thdash_{\SEQCALCD^{cut}}\Delta$ with $d$ proof steps and $n$ applications of
  rule $\seqcut$ there exists an alternative cut-free derivation
  $\cD'\colon\thdash_{\SEQCALCD}\Delta$ with maximally $d+nk$ proof steps.
\end{cor}

\section{The Extensionality Axioms are Cut-Strong}\label{sec:prominent-cut-strong}
We have shown comprehension axioms can be cut-strong (cf. Example~\ref{ex4}).  Further
prominent examples of cut-strong formulae are the Boolean and functional extensionality
axioms. The Boolean extensionality axiom (abbreviated as $\axiomb$ in the remainder) is
\[\alldot{A\abtypebool}\alldot{B\abtypebool} (A\follof B)\implies
A\Leibeq^\typebool B\] The infinitely many functional extensionality axioms (abbreviated
as $\axiomf$) are parameterized over ${\typea,\typeb}\in\Types$.
\[\alldot{F\abtype{\typea\ar\typeb}}\alldot{G\abtype{\typea\ar\typeb}}
        (\alldot{X\abtypea} FX\Leibeq^\typeb GX)\implies
        F\Leibeq^{\typea\ar\typeb} G
\] 

These axioms usually have to be added to higher-order calculi to reach Henkin
completeness, i.e. completeness with respect to model class $\MODFB$.  For
example, Huet's constrained resolution approach as presented
in~\cite{Huet:amott73} is not Henkin complete without adding extensionality
axioms. The need for adding Boolean extensionality to this calculus is actually
illustrated by the set of unit literals $\Phi\deq\{a,b,(qa),\neg (q b)\}$ from
Example~\ref{ex:nosatb}. As the reader may easily check, this clause set $\Phi$,
which is inconsistent for Henkin semantics, cannot be proven by Huet's system
without, e.g., adding the Boolean extensionality axiom.  By relying on results
in~\cite{Andrews:ritt71}, Huet essentially shows completeness with respect to
model class $\MODD$ as opposed to Henkin semantics.

We will now investigate whether adding the extensionality axioms to a machine-oriented
calculus in order to obtain Henkin completeness is a suitable option.

\begin{thm}\label{thm:cut-strong-bool} The Boolean extensionality axiom
  $\axiomb$ is a $14$-cut-strong formula in $\SEQCALCD$.
\end{thm}

\proof The following derivation justifies this theorem
($a_\typebool$ is a parameter).  

{
 \[ \infer[2\times\seqpilg{a}]{\Delta*\neg\axiomb}
        {
          \infer[\seqlorl]{\Delta*\neg(\neg(a\follof a)\vee
            a\Leibeq^\typebool a)}
                {\infer[\seqneg]{\Delta*\neg\neg(a\follof
        a)}
                  {\infer*[]{\Delta*a\follof a}{\mbox{7 steps; see Lemma~\ref{lemma-length}}}}
                &
                \infer*[\mbox{3 steps; see Ex.~\ref{ex3}}]{\Delta*\neg(a\Leibeq^\typebool a)}{\deduce{}{\Delta*\bC & \Delta*\neg\bC}}  
                }
        }
 \]
}

\qed

\begin{thm}\label{thm:cut-strong-func} 
  The functional extensionality axioms $\axiomf$ are $11$-cut-strong formulae in
  $\SEQCALCD$.
\end{thm}

\proof The following derivation justifies this theorem ($f_{\typea\ar\typeb}$ is a
parameter).
{
\[
  \infer[2\times\seqpilg{f}]{\Delta*\neg\axiomf}
        {
          \infer[\seqlorl]{\Delta*\neg (\neg (\alldot{X\abtypea} fX\Leibeq^\typeb fX)\vee
        f\Leibeq^{\typea\ar\typeb} f)}
                {\infer[\seqneg]{\Delta*\neg\neg\alldot{X\abtypea} fX\Leibeq^\typeb fX}
                  {\infer[\seqpirg{a\abtypea}]
                    {\Delta*(\alldot{X\abtypea} fX\Leibeq^\typeb
  fX)}
                    {\infer*[]
                      {\Delta*fa\Leibeq^\typeb fa}
                      {\mbox{3 steps; see Lemma~\ref{lemma-length}}}}}
                &
                \infer*[\mbox{3 steps; see Ex.~\ref{ex3}}]
                       {\Delta*\neg(f\Leibeq^{\typea\ar\typeb} f)}
                       {\deduce{}{\Delta*\bC & \Delta*\neg\bC}}  
                }
        }
\]

\qed
}

In~\cite{Ben99} and~\cite{Brown2004a} we have already argued that the
extensionality principles should not be treated axiomatically in
machine-oriented higher-order calculi and there we have developed
resolution and sequent calculi in which these principles are built-in. Here
we have now developed a strong theoretical justification for this work:
Corollary~\ref{thm:cut-simulation-c} along with
Theorems~\ref{thm:cut-strong-func} and ~\ref{thm:cut-strong-bool}
tell us that adding the extensionality
principles $\axiomb$ and $\axiomf$ as axioms to a
calculus is like adding a cut rule.


 In Figure~\ref{seqcalcaxiomext} we show rules that add Boolean and
 functional extensionality in an axiomatic manner to $\SEQCALCD$. More
 precisely we add rules $\seqaxiomf$ and $\seqaxiomb$
 allowing to introduce the axioms for any sequent $\Delta$; this way we
 address the problem of the infinitely many possible instantiations
 of the type-schematic functional extensional axiom
 $\axiomf$. 

\begin{figure}[ht]
\begin{center}
\begin{textnd}
    \ibnc{\Delta *\neg\axiomf}
         {{\typea\ar\typeb}\in\Types}
         {\Delta }
         {\seqaxiomf}\hspace*{3em}
   \ianc{\Delta *\neg\axiomb}
        {\Delta }
         {\seqaxiomb}\\[-2em]
\end{textnd}
\end{center}
\caption{\label{seqcalcaxiomext} Axiomatic Extensionality Rules}
\end{figure}

 Calculus $\SEQCALCD$ enriched by the new rules
 $\seqaxiomf$ and $\seqaxiomb$ is called $\SEQCALCD^E$. Soundness of the
 the new rules is easy to verify: In \pubjslb{4.3}~we show that
 $\seqaxiomf$ and $\seqaxiomb$ are valid for Henkin models.

\subsection{Replacing the Extensionality Axioms.}
 In Figure~\ref{seqcalcext} we define alternative extensionality rules
 which correspond to those developed for resolution and sequent calculi
 in~\cite{Ben99} and~\cite{Brown2004a}.

\begin{figure}[ht]
\begin{center}
\begin{textnd} 
    \ianc{\Delta *\bnormform{(\alldot {X_\typea}\bA X\Leibeq^{\typeb}\bB X)}}
         {\Delta * (\bA\Leibeq^{\typea\ar\typeb}\bB)}
         {\seqpropf}
 \hspace*{3em}  \ibnc{\Delta *\neg\bA *\bB}
         {\Delta *\neg\bB *\bA}
         {\Delta * (\bA\Leibeq^\typebool\bB)}
         {\seqpropb}\\[-.2cm]
\end{textnd}
\end{center}
\caption{\label{seqcalcext} Proper Extensionality Rules}
\end{figure}

 Calculus $\SEQCALCD$ enriched by
 $\seqpropf$ and $\seqpropb$ is called $\SEQCALCFB^-$.  Soundness of
 $\seqpropf$ and $\seqpropb$ for Henkin semantics is again easy to show.

 Our aim is to develop a machine-oriented sequent calculus for automating
 Henkin complete proof search. We argue that for this purpose $\seqpropf$
 and $\seqpropb$ are more suitable rules than $\seqaxiomf$ and
 $\seqaxiomb$.

Our next step now is to show Henkin completeness for $\SEQCALCD^E$.  This
will be relatively easy since we can employ cut-simulation.  Then we analyze
whether calculus $\SEQCALCFB^-$ has the same deductive power as
$\SEQCALCD^E$.

First we extend Theorem~\ref{thm:accseq-acc}.  The proof is given in the Appendix.

\begin{thm}\label{thm:accseq-acc-fb}
  Let $\SEQCALC$ be a sequent calculus such that $\seqneginv$ and $\seqneg$
  are admissible.\shortskip
  \begin{enumerate}[\em(1)]
  \item If $\seqpropf$ and $\seqpir$ are admissible, then $\accseq\SEQCALC$ satisfies
    $\absf$.\shortskipb
  \item If $\seqpropb$ is admissible, then $\accseq\SEQCALC$ satisfies $\absb$.
  \end{enumerate}
\end{thm}

\begin{thm}
  The sequent calculus $\SEQCALCD^E$ is Henkin complete and the rule $\seqcut$ is
  $12$-admissible.
\end{thm}

\proof $\seqcut$ can be effectively simulated and hence
eliminated in $\SEQCALCD^E$ by combining rule $\seqaxiomf$ with the $11$-step derivation
presented in the proof of Theorem~\ref{thm:cut-strong-func}.  


Let $\accseq{\SEQCALCD^E}$ be defined as in Definition~\ref{def:accseq}.  We prove Henkin
completeness of $\SEQCALCD^E$ by showing that the class $\accseq{\SEQCALCD^E}$ is a
saturated abstract consistency class in $\ACCMODFB$. We here only analyze the crucial
conditions $\absb$, $\absf$ and $\abssat$. For the other conditions we refer to
Theorem~\ref{thm:accseq-acc}.  Note that $0$-admissibility of $\seqneginv$ and $\seqweak$
can be shown for $\SEQCALCD^E$ by a suitable induction on derivations as in
Lemma~\ref{lem:admissible-rules}.
\begin{description}
\absitem{\absf} $\seqpir$ is a rule of $\SEQCALCD^E$ and thus
  admissible. According to Theorem~\ref{thm:accseq-acc-fb} it is thus
  sufficient to ensure admissibility of rule $\seqpropf$ to show
  $\absf$. This is justified by the following derivation where $\bN\deq\bA
  \Leibeq^{\typea\ar\typeb}\bB$ and $\bM\deq(\alldot {X_\typea}\bA X
  \Leibeq^{\typeb}$ $\bnormform{\bB X)}$ (for $\beta$-normal $\bA,\bB$).\shortskip
{
\[ 
\infer[\seqaxiomf]
      {\Delta*\bA\Leibeq^{\typea\ar\typeb}\bB}
      {\infer[\seqpilg\bA,\seqpilg\bB]
        {\Delta*{\bN}*\neg\axiomf}
        {\infer[\seqlorl]
          {\Delta*{\bN}*\neg(\neg{\bM}\vee
            {\bN})}
          {\infer[\seqneg]
            {\Delta*{\bN}*\neg\neg{\bM}}
            {\infer[\seqweak]
              {\Delta*{\bN}*{\bM}}
              {\Delta*\bnormform{(\alldot {X_\typea}\bA X\Leibeq^{\typeb}\bB X)}}
              } 
            &
            \hspace{-1.5em}\infer*[]
                   {\Delta*{\bN}*\neg{\bN}}
                   {\mbox{derivable}}
          }
        }
      }
\shortskipb
\]
} 
\absitem{\absb} With a similar derivation using $\seqaxiomb$ we can show that $\seqpropb$
  is admissible. We conclude $\absb$ by Theorem~\ref{thm:accseq-acc-fb}.\shortskip
\absitem{\abssat} Since $\seqcut$ is admissible we get saturation by
Theorem~\ref{lem:cut-implies-set}. \qed
\end{description}

\noindent Does $\SEQCALCFB^-$ have the same deductive strength as
$\SEQCALCD^E$?  I.e., is $\SEQCALCFB^-$ Henkin complete? We show this
is not yet the case.




\begin{thm}
The sequent calculus $\SEQCALCFB^-$ is not complete for Henkin semantics.
\end{thm}

We illustrate the problem by a counterexample.

\begin{exa}\label{ex-incomplb}
  Consider the sequent $\Delta\deq\{\neg a,\neg b,\neg (qa),$ $ (qb)\}$
  where $a_\typebool,b_\typebool,$\linebreak 
  $q_{\typebool\ar\typebool}\in\Signat$ are
  parameters.  For any $\cM\Metaeq(\cD,\appo,\cE,\semival)\in\MODFB$,
  either $\semival(\cE(a))\Metaeq\semfalse$,
  $\semival(\cE(b))\Metaeq\semfalse$ or $\cE(a)\Metaeq\cE(b)$ by property
  $\propb$.  Hence sequent $\Delta$ is valid for every $\cM\in\MODFB$.
  However, $\thdash_{\SEQCALCFB^-}\Delta$ does not
  hold. By inspection, $\Delta$ cannot be the
  conclusion of any rule.
\end{exa}

In order to reach Henkin completeness and to show cut-elimination we thus
need to add further rules.  Our example motivates the two rules presented
in Figure~\ref{seqcalcdec}.  $\seqinitleib$ introduces Leibniz equations
such as $qa\Leibeq^\typebool qb$ as is needed in our example and $\seqdec$
realizes the required decomposition into $a\Leibeq^\typebool b$.

\begin{figure}[ht]
\[
  \ibnc{\Delta * (\bA\Leibeq^\typebool\bB)}{(\dagger)}
       {\Delta *\neg\bA *\bB}
       {\seqinitleib}\hspace*{3em}
       {\ibnc
         {\Delta * (\bA^1\Leibeq^{\typea_1}\bB^1)\;\cdots\;
          \Delta * (\bA^n\Leibeq^{\typea_n}\bB^n)}
         {(\ddagger)}
         {\Delta * (h\ov{\bA^n}\Leibeq^{\typeb} h\ov{\bB^n})}
         {\seqdec}
       }
\]

$(\dagger)\hspace{1em}\bA {\mbox{,}}\bB {\mbox{ atomic}}$\hfill $(\ddagger)\hspace{1em} {n\geq 1,\typeb\in\{\typebool,\typeind\}, h_{\ov{\typea^n}\ar\typeb}\in\Signat {\mbox{ parameter}}}$
\caption{\label{seqcalcdec} Additional Rules $\seqinitleib$ and $\seqdec$}
\end{figure}

We thus extend the sequent calculus $\SEQCALCFB^-$ to $\SEQCALCFB$ by adding the
decomposition rule $\seqdec$ and the rule $\seqinitleib$ which generally checks if two
atomic sentences of opposite polarity are provably equal (as opposed to syntactically
equal).

Is $\SEQCALCFB$ complete for Henkin semantics? We will show in the next Section that this
indeed holds (cf. Theorem~\ref{theo:fb-complete}).

With $\SEQCALC^E$ and $\SEQCALCFB$ we have thus developed two Henkin complete calculi and
both calculi are cut-free. However, as our exploration shows, ``cut-freeness'' is not a
well-chosen criterion to differentiate between their suitability for proof search
automation: $\SEQCALC^E$ inherently supports effective cut-simulation and thus
cut-freeness is meaningless.

The next claim, which is analogous to Theorem ~\ref{firstclaim}, has not been
formally proven yet. It claims that, in contrast to $\SEQCALC^E$,
the cut-freeness of $\SEQCALCFB$ is meaningful.

\begin{Claim}\label{secondclaim}
  $\seqcut$ is not $k$-admissible in $\SEQCALCFB$.
\end{Claim}

The proof idea is similar to that of Theorem ~\ref{firstclaim}, however, the two
additional rules $\seqinitleib$ and $\seqdec$ do introduce additional
technicalities which we have not fully worked out yet.
 

The criterion we propose for the analysis of calculi in impredicative
logics is ``freeness of effective cut-simulation''. The idea behind
this notion is to capture also hidden sources (such as
the extensionality axioms) where the
subformula property may break and where the cut rule may creep in through the backdoor.

\subsection{Other Rules for Other Model Classes.} 
In {\pubsekia} we developed respective complete and cut-free sequent calculi not only for
Henkin semantics but for five of the eight model
classes. 
In particular, no additional rules are required for the $\beta$, $\beta\eta$ and
$\beta\xi$ case.  Meanwhile, the $\beta\propf$ case requires additional rules allowing
$\eta$-conversion.  We do not present and analyze these cases here.

\section{Acceptability Conditions}\label{sec:acceptability}
We now turn our attention again to the existence of saturated extensions of abstract
consistency classes.

As illustrated by Example~\ref{ex:nosatb}, we need some extra abstract consistency
properties to ensure the existence of saturated extensions.  We call these extra
properties {\defins{acceptability condition}}. They actually closely correspond to
additional rules $\seqinitleib$ and $\seqdec$.

\begin{defi}[Acceptability Conditions]\label{Def:accept-acc-conds}
  Let $\acc$ be an abstract consistency class in $\ACCMODFB$.  We define the following
  properties:
  \begin{itemize}
  \absitem{\absmate} If $\bA,\bB\in\Wffclbool$ are atomic and
    $\bA,\neg\bB\in\Phi$, then $\Phi*\neg(\bA\Leibeq^\typebool\bB)\in\acc$.
  \absitem{\absdecd} If $\neg(h\ov{\bA^n}\Leibeq^{\typeb}
    h\ov{\bB^n})\in\Phi$ for some types $\typea_i$ where
    $\typeb\in\{\typebool,\typeind\}$ and $h_{\ov{\typea^n}\ar\typeb}\in\Signat$ is a parameter,
    then there is an $i$ ($1\leq i\leq n$) such that
    $\Phi*\neg(\bA^i\Leibeq^{\typea^i}\bB^i)\in\acc$.
  \end{itemize}
\end{defi}

\noindent We now replace the strong saturation condition used in
          {\pubjsla} by these acceptability conditions.

\begin{defi}[Acceptable Classes] An abstract
  consistency class $\acc\in\ACCMODFB$ is called {\defin{acceptable}} in $\ACCMODFB$ if
  it satisfies the conditions $\absmate$ and $\absdecd$.
\end{defi}

One can show a model existence theorem for acceptable abstract consistency
classes in $\ACCMODFB$ (cf.~\pubsekib{8.1}).  From this model existence
theorem, one can conclude $\SEQCALCFB$ is complete for $\MODFB$ (hence for
Henkin models) and that cut is admissible in $\SEQCALCFB$. 

\begin{thm}\label{theo:fb-complete}
The sequent calculus $\SEQCALCFB$ is complete for Henkin
semantics and the rule $\seqcut$ is admissible.
\end{thm}
\proof The
argumentation is similar to Theorem~\ref{thm:complete-beta} but here we employ the acceptability
conditions $\absmate$ and $\absdecd$.\qed

One can further show the {\bf Saturated Extension
Theorem} (cf.~\pubsekib{9.3}):

\begin{thm} There is a saturated
abstract consistency class in $\ACCMODFB$ that is an extension of all
acceptable $\acc$ in $\ACCMODFB$.
\end{thm}

Given Theorem~\ref{thm:satext-implies-cut}, one can view the Saturated Extension Theorem
as an abstract cut-elimination result.

The proof of a model existence theorem employs Hintikka sets and in the
context of studying Hintikka sets we have identified a
phenomenon related to cut-strength which we call the \defin{Impredicativity
Gap}. That is, a Hintikka set $\cH$ is saturated if any cut-strong formula
$\bA$ (e.g. a Leibniz equation $\bC\Leibeq\bD$) is in $\cH$.  Hence we can
reasonably say there is a ``gap'' between saturated and unsaturated
Hintikka sets.  Every Hintikka set is either saturated or contains no
cut-strong formulae.

\section{Conclusion}
We have shown that adding cut-strong formulae to a calculus for
an impredicative logic is like adding cut.  For machine-oriented automated
theorem proving in impredicative logics --- such as classical type theory
--- it is therefore not recommendable to naively add cut-strong axioms to
the search space.  In addition to the comprehension principle and the
functional and Boolean extensionality axioms as elaborated in this paper
the list of cut-strong axioms includes:

\begin{exa}[Other Forms of Defined Equality.]
  Formulas $\bA\Andrewseq^\alpha\bB$ are $4$-cut-strong in $\SEQCALCD$ where
  $\Andrewseq^\alpha$ is $\lamdot{X_\alpha}\lamdot{Y_\alpha} \alldot{Q_{\alpha\ar\alpha\ar
      o}} (\alldot{Z_\alpha} (Q\ Z\ Z))\implies (Q\ X\ Y)$ (cf.~\cite{Andrews02}). The
  argument is similar to Examples~\ref{ex1}-\ref{ex4}; here we the crucial step is to
  instantiate $Q$ with $\lambda X_\alpha\sdot\lambda Y_\alpha \sdot\bC$.
\end{exa}

\begin{exa}[Axiom of Induction.]
  The \emph{axiom of induction} for the naturals
  $\alldot{P_{\typeind\ar\typebool}} P 0\wedge (\alldot{X_{\typeind}}
  PX\implies P(sX))\implies\alldot{X_{\typeind}} PX$ is
  $18$-cut-strong in $\SEQCALCD$. (Other well-founded ordering axioms
  are analogous.)  The crucial step in the proof is to instantiate $P$
  with $\lamdot{X_{\typeind}} a\Leibeq^\typebool a$ for some parameter
  $a_\typebool$.
\end{exa}

\begin{exa}[Axiom of Choice.]
  {$\exdot{I_{(\typea\ar\typebool)\ar\typebool}}\alldot{Q_{\typea\ar\typebool}}
    (\exdot{X_\typea} Q X)\implies Q (I Q)$} is $7$-cut-strong in
  $\SEQCALCD$.  The crucial step is to instantiate $Q$ with
  $\lamdot{X_\typea} {\bC}$.
\end{exa}

\begin{exa}[Axiom of Description.]
  $\exdot{I_{(\typea\ar\typebool)\ar\typebool}}
  \alldot{Q_{\typea\ar\typebool}} (\exists_1 Y_\typea\sdot Q
  Y)\implies Q (I Q)$, the \emph{description axiom}
  (see~\cite{Andrews:gmae72}), where $\exists_1 Y_\typea\sdot Q Y$
  stands for $\exdot{Y_\typea} Q Y\wedge (\alldot{Z_\typea} QZ\implies
  Y\Leibeq Z)$ is $25$-cut-strong in $\SEQCALCD$.  The crucial step in
  the proof is to instantiate $Q$ with $\lamdot{X_\typea}
  a\Leibeq^\typea X$ for some parameter $a_\typea$.
\end{exa}

As we have shown in Example~\ref{ex4}, comprehension axioms can be cut-strong.  Church's
formulation of type theory (cf.~\cite{Church40}) used typed $\lambda$-calculus to build
comprehension principles into the language.  One can view Church's formulation as a first
step in the program to eliminate the need for cut-strong axioms.  For the extensionality
axioms a start has been made by the sequent calculi in this paper (and {\pubsekia}), for
resolution in~\cite{Ben99} and for sequent calculi and extensional expansion proofs
in~\cite{Brown2004a}.  The extensional systems in~\cite{Brown2004a} also provide a
complete method for using primitive equality instead of Leibniz equality.  For improving
the automation of higher-order logic our exploration thus motivates the development of
higher-order calculi which directly include reasoning principles for equality,
extensionality, induction, choice, description, etc., without using cut-strong axioms.


%

\section*{Acknowledgement}
We thank the reviewers of this paper for their useful comments and suggestions.

\bibliographystyle{plain}

\pagebreak 

\section*{Appendix}

\paragraph{\bf Proof of Lemma~\ref{lem:accseq-notin}}
\proof
  Suppose $\Phi\cup\neg\Delta\notin\accseq\SEQCALC$.
  By definition, $\thdash_{\SEQCALC} \neg\bnormform\Phi\cup\neg\neg\bnormform\Delta$
  holds.  Applying $\seqneginv$ to each member of $\bnormform\Delta$,
  we have $\thdash_{\SEQCALC} \neg\bnormform\Phi\cup\bnormform\Delta$.
\qed

\paragraph{\bf Proof of Theorem~\ref{thm:accseq-acc}:} 
\proof 
  We prove $\accseq\SEQCALC$ is closed under subsets and
  satisfies $\absc$, $\absneg$, $\absor$, $\absand$ and $\absbeta$.  The remaining
  conditions are proven analogously.
  
  Suppose $\Phi\in\accseq\SEQCALC$, If $\Phi_0\subseteq\Phi$ and
  $\Phi_0\notin\accseq\SEQCALC$, then $\thdash_\SEQCALC\neg\bnormform{\Phi_0}$ and so
  $\thdash_\SEQCALC\neg\bnormform{\Phi}$ by admissibility of $\seqweak$.  Hence
  $\accseq\SEQCALC$ is closed under subsets.

  Suppose $\Phi\in\accseq\SEQCALC$ and $\bA,\neg\bA\in\Phi$
  where $\bA$ is atomic.  By admissibility of $\seqinit$, $\thdash_\SEQCALC\neg\bnormform{\Phi}*\bnormform\bA$
  since $\neg\bnormform\bA\in\neg\bnormform\Phi$.  By admissibility of $\seqneg$,
  $\thdash_\SEQCALC\neg\bnormform\Phi$ since $\neg\neg\bnormform\bA\in\neg\bnormform\Phi$,
  contradicting $\Phi\in\accseq\SEQCALC$.
  Thus $\absc$ holds.

  Suppose $\Phi\in\accseq\SEQCALC$, $\neg\neg\bA\in\Phi$ and
  $\Phi*\bA\notin\accseq\SEQCALC$.  Hence $\thdash_\SEQCALC
  \neg\bnormform\Phi*\neg\bnormform\bA$ and so $\thdash_\SEQCALC
  \neg\bnormform\Phi*\neg\neg\neg\bnormform\bA$ by admissibility of
  $\seqneg$.  Since $\neg\neg\bA\in\Phi$, we know $\neg\bnormform\Phi$ is
  equal to $\neg\bnormform\Phi*\neg\neg\neg\bnormform\bA$.  Hence
  $\thdash_\SEQCALC \neg\bnormform\Phi$, contradicting
  $\Phi\in\accseq\SEQCALC$.  Thus $\absneg$ holds.  

  Suppose $\Phi\in\accseq\SEQCALC$, $(\bA\lor\bB)\in\Phi$,
  $\Phi*\bA\notin\accseq\SEQCALC$ and $\Phi*\bB\notin\accseq\SEQCALC$.
  Hence $\thdash_\SEQCALC \neg\bnormform\Phi*\neg\bnormform\bA$ and
  $\thdash_\SEQCALC \neg\bnormform\Phi*\neg\bnormform\bB$.  Applying
  $\seqlorl$, we have $\thdash_\SEQCALC \neg\bnormform\Phi$ since
  $\neg\bnormform{(\bA\lor\bB)}\in\neg\bnormform\Phi$, contradicting
  $\Phi\in\accseq\SEQCALC$.  Thus $\absor$ holds.

  By a similar argument, admissibility of $\seqpil$ implies $\absforall$.

  Suppose $\Phi\in\accseq\SEQCALC$, $\neg(\bA\lor\bB)\in\Phi$ and
  $\Phi*\neg\bA*\neg\bB\notin\accseq\SEQCALC$.  By
  Lemma~\ref{lem:accseq-notin}, $\thdash_\SEQCALC
  \neg\bnormform\Phi*\bnormform\bA*\bnormform\bB$.  Applying $\seqlorr$, we
  have $\thdash_\SEQCALC \neg\bnormform\Phi*\bnormform{(\bA\lor\bB)}$.
  Applying $\seqneg$, we have $\thdash_\SEQCALC \neg\bnormform\Phi$ since
  $\neg(\bA\lor\bB)\in\Phi$, contradicting $\Phi\in\accseq\SEQCALC$.  Thus
  $\absand$ holds.
 
  By a similar argument, admissibility of $\seqpir$, $\seqneginv$ and
  $\seqneg$ imply $\absexists$.

  Suppose $\Phi\in\accseq\SEQCALC$, $\bA\in\Phi$, $\bA\eqb\bB$
  and $\Phi*\bB\notin\accseq\SEQCALC$.
  Hence $\thdash_\SEQCALC\neg\bnormform\Phi*\neg\bnormform\bB$,
  contradicting $\bnormform\bA\in\bnormform\Phi$ and $\Phi\in\accseq\SEQCALC$.
  Thus $\absbeta$ holds.
\qed

\paragraph{\bf Proof of Theorem~\ref{lem:cut-implies-set}:}
  \proof
  Suppose $\seqcut$ is admissible, $\Phi\in\accseq\SEQCALC$, $\bA\in\cWffbool$,
  $\Phi*\bA\notin\accseq\SEQCALC$ and $\Phi*\neg\bA\notin\accseq\SEQCALC$.
  Hence $\thdash_\SEQCALC\neg\bnormform\Phi*\neg\bnormform\bA$ and
  $\thdash_\SEQCALC\neg\bnormform\Phi*\neg\neg\bnormform\bA$.  Using $\seqcut$, we have
  $\thdash_\SEQCALC\neg\bnormform\Phi$, contradicting $\Phi\in\accseq\SEQCALC$.

  Suppose $\accseq\SEQCALC$ is saturated, $\thdash_\SEQCALC\Delta*\bC$ and
  $\thdash_\SEQCALC\Delta*\neg\bC$ hold but $\thdash_\SEQCALC\Delta$ does
  not.  Applying $\seqneg$ to every member of $\Delta$ and to $\bC$ we have
  $\thdash_\SEQCALC\neg\neg\Delta*\neg\neg\bC$ and
  $\thdash_\SEQCALC\neg\neg\Delta*\neg\bC$.  By
  Lemma~\ref{lem:accseq-notin}, we know $\neg\Delta\in\accseq\SEQCALC$.  By
  saturation, we must have $\neg\Delta*\bC\in\accseq\SEQCALC$ or
  $\neg\Delta*\neg\bC\in\accseq\SEQCALC$.  The first case contradicts
  $\thdash_\SEQCALC\neg\neg\Delta*\neg\bC$ while the second case
  contradicts $\thdash_\SEQCALC\neg\neg\Delta*\neg\neg\bC$.  \qed

\paragraph{\bf Proof of Lemma~\ref{thm:satext-implies-cut}:} 
\proof Suppose $\acc'\in\ACCstar$ is a saturated extension of $\accseq\SEQCALC$.  Assume
$\thdash_\SEQCALC\Delta*\bC$ and $\thdash_\SEQCALC\Delta*\neg\bC$ hold and
$\thdash_\SEQCALC\Delta$ does not.  By Lemma~\ref{lem:accseq-notin}, we know
$\neg\Delta\in\accseq\SEQCALC$.  Since $\neg\Delta$ is finite (hence sufficiently
$\Signat$-pure), $\neg\Delta\in\acc'$.  By the model existence theorem for saturated
abstract consistency classes (cf. Theorem~\pubjslb{6.34}), there is a model
$\cM\in\MODALL$ such that $\cM\models\neg\Delta$.  By soundness of $\accseq\SEQCALC$, we
know both $\Delta*\bC$ and $\Delta*\neg\bC$ must be valid in $\cM$.  Since
$\cM\models\neg\Delta$, we must have $\cM\models\bC$ and $\cM\models\neg\bC$, a
contradiction.  \qed

\paragraph{\bf Proof of Lemma~\ref{lem:admissible-rules}:} 
  \proof
  We can argue $0$-admissibility of $\seqneginv$ and $\seqweak$ by
  induction on derivations.
  We use the notation $\thdash_{\SEQCALCD}^n\Delta$ to indicate
  there is a derivation with size at most $n$ of $\Delta$.
  For negation inversion, we need to show $\thdash_{\SEQCALCD}^n\Delta*\bA$
  whenever $\thdash_{\SEQCALCD}^n\Delta*\neg\neg\bA$.
  First assume $\neg\neg\bA$ is a principal formula of the last rule
  applied.  This is only possible if the last rule is $\seqneg$.
  Examining $\seqneg$,
  we have either $\thdash_{\SEQCALCD}^{n-1}\Delta*\bA$
  or  $\thdash_{\SEQCALCD}^{n-1}\Delta*\neg\neg\bA*\bA$.
  In the first case, we are done.  Otherwise, we apply
  the induction hypothesis to $\thdash_{\SEQCALCD}^{n-1}\Delta*\neg\neg\bA*\bA$
  and obtain $\thdash_{\SEQCALCD}^{n-1}\Delta*\bA$ as desired.
  Next assume $\neg\neg\bA$ is not a principal formula of
  the last rule.  In this case, the application of rule $r$ concludes
  $\thdash_{\SEQCALCD}^{n} (\Delta'*\neg\neg\bA)\cup\Delta_0$
  from 
  $\thdash_{\SEQCALCD}^{n^i} (\Delta'*\neg\neg\bA)\cup\Delta_i$
  (with $1\leq i\leq m$) where $\Delta_0$ contains the principal
  formulae of the rule application (a singleton unless the rule is $\seqinit$)
  and
  $n^1+\cdots+n^m\leq n-1$.
  Applying the inductive hypothesis, we have
  $\thdash_{\SEQCALCD}^{n^i} (\Delta'*\bA)\cup\Delta_i$ for $1\leq i\leq m$.
  Applying rule $r$ we have $\thdash_{\SEQCALCD}^{n} (\Delta'*\bA)\cup\Delta_0$.
  (For the case where $r$ is $\seqpir$ we use the fact that
  the same parameters occur in $\bA$ and $\neg\neg\bA$.)

  To prove $0$-admissibility of weakening, we generalize the
  statement to include a parameter renaming (to handle the $\seqpir$ rule).
  A parameter renaming $\theta$ is a well-typed map from parameters to parameters
  extended to operate on arbitrary terms.  Note that if $\bA$ is $\beta$-normal,
  then $\theta(\bA)$ is also $\beta$-normal.    Also,
  if $\bA$ is atomic, then $\theta(\bA)$ is atomic.
  We prove for any $n$, $\Delta$,
  $\Delta'$
  and parameter renaming $\theta$, if $\thdash_{\SEQCALCD}^n\Delta$
  and $\theta(\bA)\in\Delta'$ for every $\bA\in\Delta$,
  then $\thdash_{\SEQCALCD}^n\Delta'$.
  Applying this with the identity parameter renaming $\theta$,
  we have $0$-admissibility of $\seqweak$.

  Suppose $\thdash_{\SEQCALCD}^n\Delta$ and $\theta(\bA)\in\Delta'$ for every $\bA\in\Delta$.
  First, assume the last rule application is $\seqpir$
  with principal formula $(\Pi^\typea\bG)\in\Delta$.
  In this case we know 
  $\thdash_{\SEQCALCD}^{n-1} \Delta_0*\bnormform{(\bG c_\typea)}$
  where $\Delta_0*(\Pi\bG)$ is $\Delta$
  and $c$ does not occur in any sentence in $\Delta$.
  Choose a parameter $d_\typea$ such that $d$ does not occur in
  any sentence in $\Delta'$.
  Let $\theta'$ be the parameter renaming given by $\theta'(c)\deq d$
  and $\theta'(w)\deq\theta(w)$ for parameters $w$ other than $c$.
  Let $\Delta''$ be $\Delta'*\bnormform{(\theta(\bG) d)}$.
  For each $\bA\in\Delta_0\subseteq\Delta$, we know $\theta'(\bA)\Metaeq\theta(\bA)\in\Delta'\subseteq\Delta''$
  (since $c$ does not occur in any sentence in $\Delta$).
  Also, since $c$ does not occur in $\bG$, $\theta'(\bnormform{(\bG c)})\Metaeq \bnormform{(\theta(\bG) d)}\in\Delta''$.
  Hence we can apply the induction hypothesis with $n-1$,
  $\Delta_0*\bnormform{(\bG c)}$,
  $\Delta''$ and $\theta'$ to conclude
  $\thdash_{\SEQCALCD}^{n-1} \Delta'*\bnormform{(\theta(\bG) d_\typea)}$.
  Since $d$ does not occur in $\Delta'$ and $\theta(\Pi\bG)\in\Delta'$, we can apply $\seqpirg{d}$
  to conclude $\thdash_{\SEQCALCD}^{n} \Delta'$.

  Next, assume the last rule applied is $\seqpil$.
  Hence
  $\thdash_{\SEQCALCD}^{n-1} \Delta_0*\neg\bnormform{(\bG \bC)}$
  where $\Delta_0*\neg(\Pi\bG)$ is $\Delta$.
  We apply the induction hypothesis with $n-1$,
  $\Delta_0*\bnormform{(\bG \bC)}$, $\Delta'*\neg\bnormform{(\theta(\bG\bC))}$
  and $\theta$ to conclude
  $\thdash_{\SEQCALCD}^{n-1} \Delta'*\neg\bnormform{(\theta(\bG\bC))}$.
  Applying the rule $\seqpilg{\theta(\bC)}$, we obtain
  $\thdash_{\SEQCALCD}^{n-1} \Delta'$ as desired.
  (Note that $\theta(\neg\Pi\bG)\in\Delta'$.)

  Finally, assume the last rule application is not $\seqpir$ and not $\seqpil$.
  Let $r$ be the last rule applied.
  The rule $r$ concludes $\thdash_{\SEQCALCD}^{n} \Delta$
  from
  $\thdash_{\SEQCALCD}^{n^i} \Delta_0\cup\Delta_i$
  where $\Delta_0\subseteq\Delta$, $1\leq i\leq m$ and $n^1+\cdots+n^m\leq n-1$.
  For each $i$,
  we can apply the induction hypothesis with $n^i$, 
  $\Delta_0\cup\Delta_i$,
  $\Delta'\cup\{\theta(\bA)|\bA\in\Delta_i\}$ and $\theta$
  to conclude $\thdash_{\SEQCALCD}^{n^i} \Delta'\cup\{\theta(\bA) | \bA\in\Delta_i\}$.
  Applying the same rule $r$ we conclude
  $\thdash_{\SEQCALCD}^{n} \Delta'$.
  \qed

\paragraph{\bf Proof of Theorem~\ref{thm:accseq-acc-fb}:}
  \proof
  Assume the rules $\seqpropf$ and $\seqpir$ are admissible.  If
  $\neg(\bG\Leibeq^{\typea\ar\typeb}\bH)\in\Phi$ and
  $\thdash_\SEQCALC\neg\bnormform\Phi*\bnormform{(\bG w\Leibeq^{\typeb}\bH
  w)}$ (with $w_\typea$ new) holds, then we can show
  $\thdash_\SEQCALC\neg\bnormform\Phi$ holds using $\seqpirg{w}$ and
  $\seqpropf$.

  Assume the rule $\seqpropb$ is admissible.  Suppose
  $\Phi\in\accseq\SEQCALC$, $\neg(\bA\Leibeq^\typebool\bB)\in\Phi$, $\Phi *
  \bA *\neg\bB\notin\accseq\SEQCALC$ and $\Phi * \neg\bA
  *\bB\notin\accseq\SEQCALC$.  By Lemma~\ref{lem:accseq-notin},
  $\thdash_\SEQCALC\neg\bnormform\Phi*\neg\bnormform\bA*\bnormform\bB$ and
  $\thdash_\SEQCALC\neg\bnormform\Phi*\bnormform\bA*\neg\bnormform\bB$.
  Applying $\seqpropb$,
  $\thdash_\SEQCALC\neg\bnormform\Phi*\bnormform{(\bA\Leibeq^\typebool\bB)}$.
  Applying $\seqneg$, $\thdash_\SEQCALC\neg\bnormform\Phi$ since
  $\neg(\bA\Leibeq^\typebool\bB)\in\Phi$, contradicting
  $\Phi\in\accseq\SEQCALC$.  Thus $\absb$ holds.  \qed

\end{document}